# Spatial Microclimatic Characterization of a Parisian "Oasis" Schoolyard

Ghid Karam[1, 2], Maïlys Chanial[1, 3], Sophie Parison[1, 2], Martin Hendel[1, 2*], Laurent Royon[1]

[1] Université Paris Cité, CNRS, LIED, UMR 8236, F-75013, Paris, France
[2] Univ Gustave Eiffel, ESIEE Paris, département SEN, F-93162, Noisy-le-Grand, France
[3] Paris City Hall, Paris, France

*Corresponding email: martin.hendel@u-paris.fr

Keywords: Mobile measurements, Microclimate, GIS, Heat stress, schoolyards

## SUMMARY

In the aftermath of the 2003 heatwave, and with growing concern over climate change, Paris City Hall has been implementing several heat mitigation strategies. One of these is the OASIS Schoolyard Strategy which aims to transform Parisian schoolyards into cool islands. Within this framework, the EU-funded ERDF UIA OASIS Project aims to study the transformation of ten schoolyards, including an evaluation of their microclimatic performance.

The present article presents case study results from one schoolyard using GIS data and fixed and mobile microclimatic measurements. An analysis method for mobile measurement data is proposed, tested and discussed on the basis of this case study.

## NOMENCLATURE

BACI: Before After Control Impact
ERDF: European Regional Development Fund
LCZ: Local Climate Zone
MRT: Mean Radiant Temperature
UIA: Urban Initiative Action
UCP: Urban Cooling Potential
UTCI: Universal Thermal Climate Index

## INTRODUCTION

As a result of climate change, many cities including Paris are expected to face a strong increase in heat-wave frequency and intensity by the end of the 21st Century, with certain regions facing the risk of becoming uninhabitable at least part of the year [1]. In Paris, the frequency of heat-waves is predicted to rise from an average of one day per year to 14-26 days per year with temperatures reaching up to 50°C [2].

One of the focus points of the Paris Resilience Strategy is adapting Parisian schoolyards to heatwaves, aiming to create a network of urban cool islands that would benefit surrounding neighborhoods. To date, schoolyards have a high impervious fraction, generally paved with small-aggregate asphalt concrete and offering little vegetation apart from varying numbers of trees. The new OASIS schoolyard design aims to promote a cooler play setting for the children, with increased vegetation fractions and unsealed soils and the use of nature-based solutions. The EU-funded ERDF UIA OASIS project aims to evaluate the renovation of ten pilot schoolyards (*Figure 1*), specifically their microclimatic and thermal performance.

The evaluation of the microclimatic performance of the "cool" renovation of an urban site is generally conducted on the basis of modelling approaches and simulated case studies [3], [4] and/or in-field monitoring of microclimatic parameters using mobile [5] or fixed measurements [6].

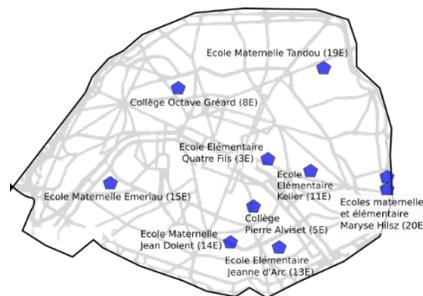

*Figure 1: The 10 pilot ERDF UIA OASIS schoolyards. Credit: City of Paris*

While fixed measurements offer a temporally continuous dataset that can be statistically analyzed, their results are only relevant to their immediate vicinity given the high spatial variability of microclimatic conditions at the pedestrian (or nano-) scale, specifically when studying heat stress. Mobile measurements offer a more spatially continuous dataset, however they only offer a limited number of observations in time for each point in space. It is therefore difficult to judge how representative or robust mobile measurements are. In addition, mobile measurements are not instantaneous and therefore the temporal drift between the first and last measurement points may be significant. Finally, comparing measurements obtained several days, weeks, months or even years apart makes comparing observations before and after a site's renovation even more difficult.

The current paper presents a case study analysis of pedestrian heat stress following the cool renovation of one of the ten pilot schoolyards: Maryse Hilsz kindergarten. Three different evaluation tools are used and compared: GIS analysis as well as mobile and fixed measurements. An analysis methodology for mobile measurement data is proposed, tested and discussed. It is compared with fixed measurement data as well as preliminary GIS data.

**METHODS**

The present paper focuses on the Maryse Hilsz kindergarten schoolyard. Fixed and mobile weather measurements are conducted to assess the microclimatic impacts of the transformation.

**Case Study Metadata**

The kindergarten Maryse Hilsz, adjacent to the Maryse Hilsz elementary school, is located in the 20[th] Arrondissement of Paris. The schoolyards of the kindergarten and elementary school are located in an area of LCZ class 5 [7]. Both schoolyards are equipped with a fixed weather station paired with a control station located 200 m outside of the schoolyards. These fixed weather stations record the microclimatic parameters listed in Table 1 with a 1-minute timestep, providing temporally continuous monitoring. This dataset is analysed following a BACI design to determine the microclimatic performance of the schoolyard renovations [8].

In addition to fixed weather stations, mobile measurements were carried out in summer before (2019) and after (2021) refurbishment works, during warm and radiative days, i.e. of Pasquill stability class A or A-B [9], presenting low wind speeds and clear skies (cover ⩽ 3 oktas), with maximum and minimum temperatures above 25° C and 16° C, respectively.

*Table 1: Fixed and mobile weather station instrumentation*

| Parameter | Instrument | Height | Uncertainty |
|---|---|---|---|
| Air Temperature, $T_a$ | Sheltered Pt 100 | 1.5/4 m | 0.1°C |
| Relative humidity, RH | Sheltered capacitive hygrometer | 1.5/4m | 1.5% RH |
| Black globe temperature, $T_g$ | Black globe Pt 100 – ISO 7726 | 1.5 m | 0.15°C |
| Wind speed, $v$ | 2D ultrasonic anemometer (fixed) | 4 m | 2% |
|  | Hot wire anemometer (mobile) | 1.5 m | 0.3 m/s |
| Net radiation, $R_n$ | Net radiometer with thermopile (fixed) | 4 m | 5% daily |

Mobile measurements are taken at 1.5 m height every 15 seconds. The white and black dots in Figure 2 illustrate the mobile measurement points within the schoolyard. The measurement points are representative of the different environments: full sun, shade, and proximity to vegetation. Following renovation works, certain measurements had to be displaced, for example due to the construction of hills and other amenities in the schoolyard, not planned at the time of the first measurement campaign. Every measurement point requires 10 to 20 minutes to allow the black globe thermometer to stabilize [10], depending on whether or not there is a change in sunlight conditions (shaded to sunny or vice versa). All measurements are carried out between 12 pm and 4 pm. Black globe temperature is used to derive mean radiant temperature (MRT) following ASHRAE Standards [11]. Collected data is used to evaluate pedestrian heat stress using UTCI [12].

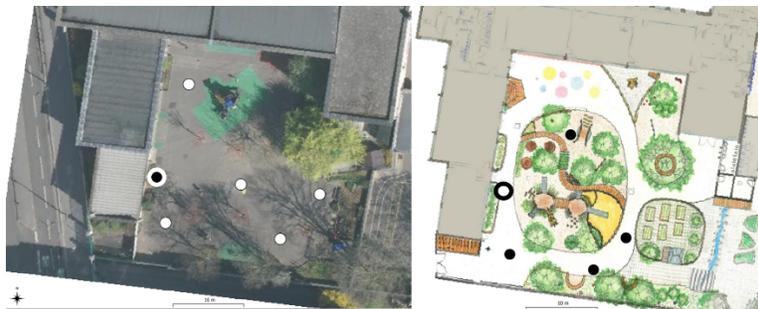

*Figure 2: Mobile measurement points at Maryse Hilsz kindergarten schoolyard before (left) and after (right) renovation. The circled point represents the fixed weather station*

**Mobile Measurement Analysis Methodology**

Despite filtering out weather conditions by selecting radiative days, a temporal correction is necessary to synchronize data obtained during mobile traverses carried out over several hours in order to take into account the evolution of the measured parameters during the measurement period [5], [13]. Similarly, as the microclimatic measurements take place up to one year apart, it is necessary to account for the differences in weather conditions between two mobile measurement campaigns. Matching mobile measurements with a fixed weather station serving as a control makes it possible to account for short- and long-term variations in weather conditions.

To this aim, we focus on the departure in pedestrian heat stress observed at a given point of the studied site, i.e. the schoolyard, from reference conditions representative of a shaded and sheltered location derived from control station measurements [14]. This parameter, named UTCI offset hereafter, is obtained following equation (1):

$$\Delta T_{UTCI,i} = T_{UTCI,i}^{mobile} - T_{UTCI,i}^{ref}\left(T_{air,i}^{control}, RH_i^{control}, T_{mrt}^{ref}, v^{ref}\right) \quad (1)$$

where UTCI offset $\Delta T_{UTCI,i}$ is the difference between the UTCI-eq temperature measured at point $i$ with mobile measurements, i.e. $T_{UTCI,i}^{mobile}$, and a reference UTCI-eq temperature $T_{UTCI}^{ref}$. $T_{UTCI}^{ref}$ is obtained using control station air temperature and humidity recorded at the same time as mobile measurements are conducted at point $i$, while MRT and wind speed are set to $T_{mrt}^{ref} = T_{air,i}^{control}$ and $v^{ref} = 0.5 \, m/s$, respectively. These reference conditions correspond to those one may expect in a shaded and sheltered location such as a courtyard, where wind speed is low and MRT follows air temperature. These reference conditions are defined such as to improve the comparability of mobile measurements between different study sites using different control stations.

Compensating for short- and long-term temporal changes in this manner is achieved if the air temperature, humidity and MRT offsets can be considered constant throughout the measurement period. While the air temperature and humidity offsets may be constant throughout the day if the control station is sufficiently representative of the study site, there is no reason for this to be true for the MRT offset. It is therefore important for mobile measurements to be conducted such as to preserve similar insolation conditions as much as possible between campaigns. Keeping the same period of the year and of the day are crucial to this effort. Using fixed station measurements, the validity of the method will be tested hereafter.

**GIS Modelling**

The ERDF OASIS project were selected following a preliminary GIS analysis that was based on morphological data, vegetative cover, and solar irradiance to calculate an urban cooling potential (UCP) indicator, described in previous work by the authors [15], [16]. UCP is an adimensional number between 0 and 1, with 1 representing an area offering high pedestrian heat stress, e.g. a bare parking lot with a pavement albedo of 0, and 0 one providing cooling from dense shade and high evapotranspiration, e.g. a dense forest.

Figure 3 shows the UCP indicator computed before and after refurbishments. Areas with high UCP are expected to present high pedestrian heat stress, while those with low UCP are expected to offer lower heat stress conditions. Field measurements provided surface albedo values. Overall, UCP is reduced after site renovation, thanks to the addition of vegetation, higher albedo pavement and soil greening and unsealing. In Figure 3, blue areas are associated with dense vegetation and tree shading, while red areas correspond to high sun-exposure asphalt pavements.

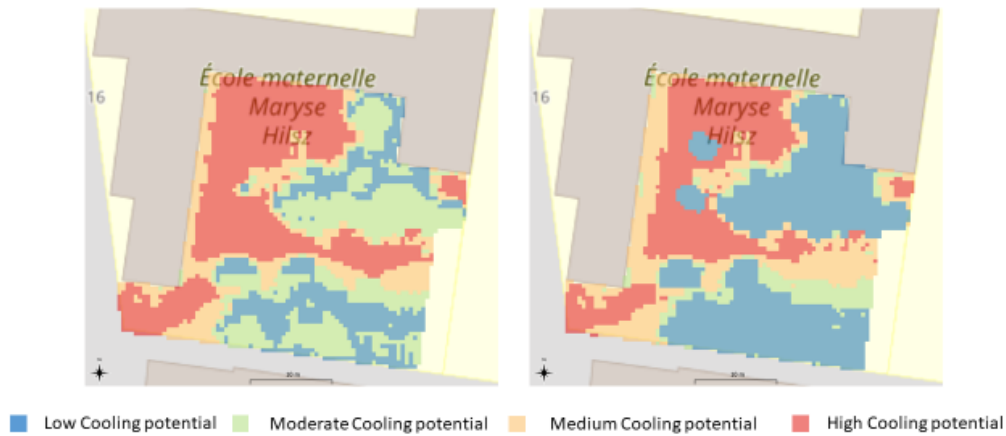

Figure 3: Hilsz Kindergarten schoolyard Heat map before (left) and after (right) retrofitting

**RESULTS AND DISCUSSION**

**Mobile Measurement Methodology**

Mobile measurements were carried out before renovation on July 25$^{th}$, 2019, and afterwards on June 15$^{th}$, 2021 between 3:30 and 5 pm local time (UTC+2). Figure 4 illustrates the microclimatic parameters that were recorded throughout the day at the fixed case station with 5-minute smoothing.

As can be seen from the top of Figure 4, the applied weather condition filtering provides for similar insolation conditions between both measurement campaigns as confirmed by net radiation and MRT. From the bottom graphs, it is apparent that the offsets are more or less constant throughout the mobile measurement period, although UTCI offset is variable on June 15$^{th}$ 2021 due to more variable insolation conditions at that moment of the day compared with 2019 measurements.

As displayed in Figure 4, microclimatic offsets are most stable during the evening and night, allowing accurate mobile measurement with little to no temporal drift following the proposed methodology. However, these periods do not allow evaluation of pedestrian heat stress during daytime, which is particularly important for spaces such as schoolyards used primarily during the day. Daytime can therefore not be discarded. From this dataset, UTCI offset can be considered constant or nearly so during the day on July 25$^{th}$ 2019 from 10 am to 3 pm within an amplitude of 2°C. On June 15$^{th}$ 2021, conditions are more variable, with UTCI-offset presenting a steady 2°C decline from 11 am to 4:30 pm and similar fluctuation amplitude.

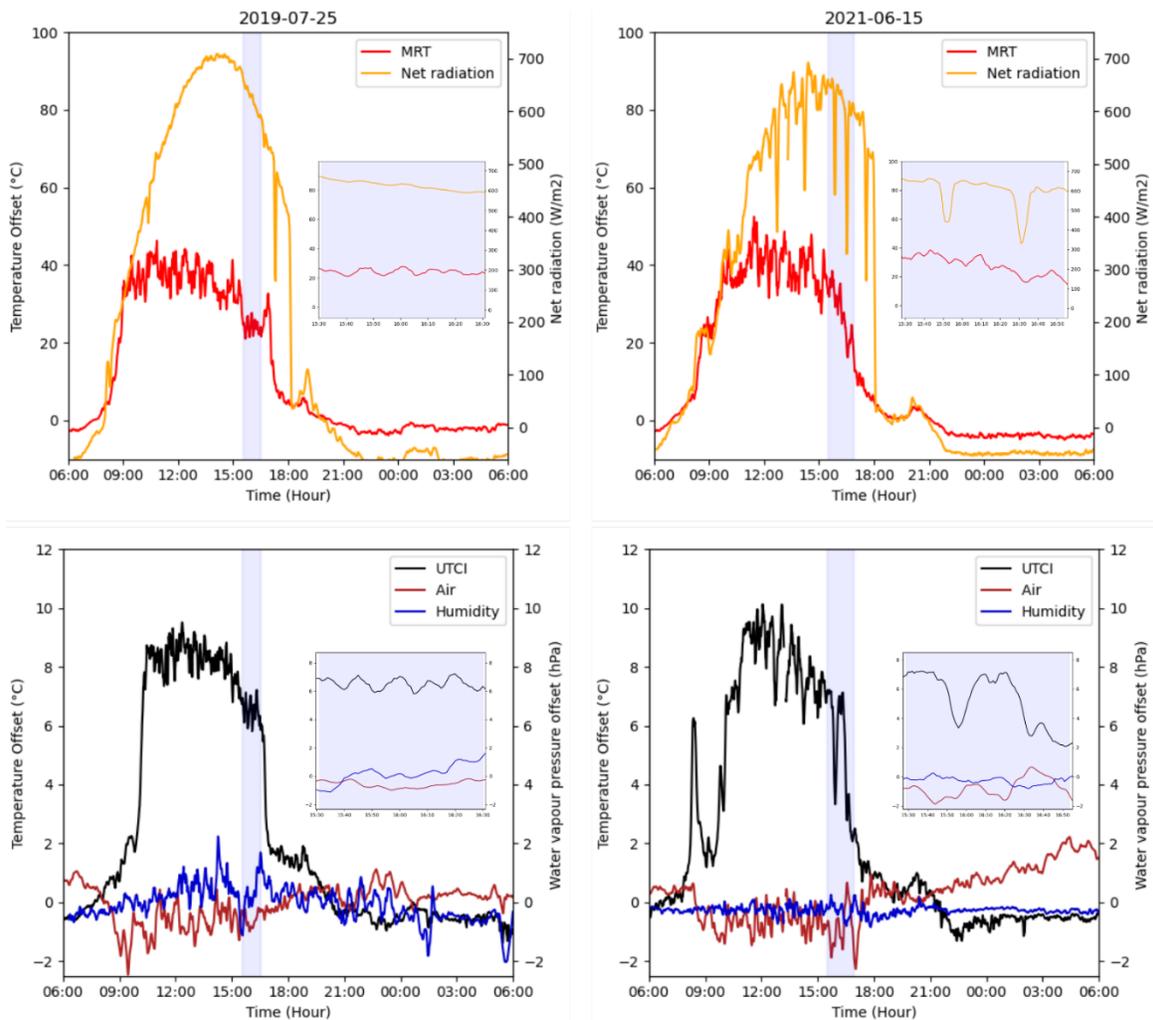

*Figure 4: Mean radiant temperature and net radiation offset (top) and UTCI-eq, air temperature and water vapour pressure offset (bottom) at the case station before (25/07/2019) and after (15/06/2021) schoolyard renovation. The mobile measurement period is highlighted by the blue shaded area and is blown out in the insets.*

In the case of a mobile measurement period that does not exceed two hours, microclimatic parameters fluctuate with a magnitude smaller than 1°C. This uncertainty may be sufficient in most cases when evaluating cooling performance on pedestrian heat stress. Further analyses will help confirm the validity of these assumptions over the course of the three years during which measurements were performed.

In addition to a control station located outside the study site, it is recommended to include a fixed measurement point where monitoring occurs continuously throughout the mobile measurement campaign. This will help verify the validity of the assumption that UTCI-offset is constant throughout the mobile measurement period.

**Pedestrian Heat Stress Map vs. Urban Cooling Potential**

Figure 5 displays a scatterplot of mobile measurements as a function of the point's estimated cooling potential.

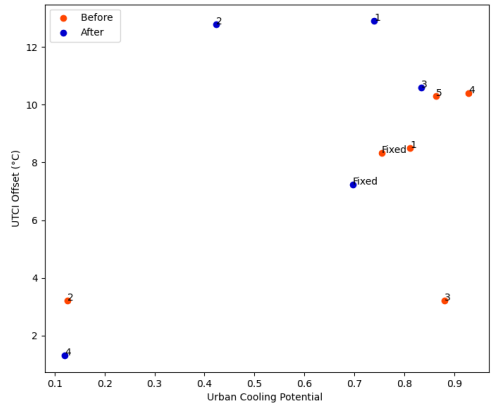

*Figure 5: UTCI offset values vs Cooling potential indicator values*

Initial observations suggest that UTCI offsets are mostly higher in high UCP areas and low ones in low UCP areas. This is consistent with the design goal of UCP and the fact that sun exposure and lack of vegetation or shading increases pedestrian heat stress. Further data points from the other studied schoolyards will help confirm these preliminary findings.

**CONCLUSIONS**

A mobile measurement methodology was proposed for use in evaluating the performance of cooling techniques applied during the renovation of an outdoor location with the aim of improving pedestrian heat stress. A fixed measurement point is necessary to identify any time drift during mobile measurements.

GIS analysis suggests a correlation between high UTCI offsets and high urban cooling potential indicator values. Such a preliminary analysis may provide useful information for designers and decision-makers prior to conducting measurements or simulations. If the validity of this analysis is confirmed, the urban cooling potential indicator might be used as a nano-scale urban descriptor as a complement to meso-scale LCZ classification in microclimate oriented urban morphological analysis.

These analyses will be conducted using the 10 pilot schoolyards to confirm these preliminary findings and help validate the applicability and validity of the proposed methodology which requires fixed control station measurements, but not fixed onsite weather measurements.

**ACKNOWLEDGEMENTS**

This research was funded by the ERDF UIA03-344 OASIS project.